\documentclass[pra,reprint,twocolumn,showpacs,superscriptaddress]{revtex4}

\usepackage{graphicx}
\usepackage{amsmath}
\usepackage{color}
\usepackage{amssymb}
\usepackage{bm}

\newcommand{\gtwofunction}{$g^{(2)}(0)$-function}
\newcommand{\gtwo}{g^{(2)}(0)}

\newcommand{\dt}{\partial_t}
\newcommand{\cimag}[1]{\mathrm{Im}\big(#1\big)}
\newcommand{\creal}[1]{\mathrm{Re}\big(#1\big)}
\newcommand{\M}{\mathrm{M}}
\newcommand{\Mi}{\mathrm{M}_i}
\newcommand{\Msi}{\mathrm{M}_i^*}
\newcommand{\I}{i}
\newcommand{\Ms}{\mathrm{M}^*}
\newcommand{\NQD}{\mathrm{N}_{\mathrm{QD}}}
\newcommand{\erw}[1]{\big\langle #1 \big\rangle}
\newcommand{\erwc}[1]{\big\langle #1 \big\rangle^c}
\newcommand{\erwt}[1]{\big\langle #1 \big\rangle_t}
\newcommand{\Gq}{\mathrm{G}_{\mathrm{q}}}
\newcommand{\Gsq}{\mathrm{G}^*_{\mathrm{q}}}
\newcommand{\Gsqs}{\mathrm{G}^*_{\mathrm{q'}}}
\newcommand{\Gqs}{\mathrm{G}_{\mathrm{q'}}}
\newcommand{\sumq}{\sum\limits_{\mathrm{q}}}
\newcommand{\sumqs}{\sum\limits_{\mathrm{q'}}}
\newcommand{\sumi}{\sum\limits_{\mathrm{i}}}
\newcommand{\omnull}{\omega_0}
\newcommand{\omq}{\omega_{\mathrm{q}}}
\newcommand{\gammaPD}{\gamma_{\mathrm{PD}}}
\newcommand{\DELTAomqnull}{\Delta^{\mathrm{q}}_{\mathrm{0}}}
\newcommand{\DELTAcvq}{\Delta^{\mathrm{cv}}_{\mathrm{q}}}
\newcommand{\DELTAomqomqs}{\Delta^{\mathrm{q}}_{\mathrm{q'}}}
\newcommand{\kappaext}{\kappa_{\mathrm{ext}}}
\newcommand{\kappah}{\kappa_{\mathrm{h}}}
\newcommand{\GAMMAextPD}{\gammaPD + \kappaext}
\newcommand{\avdag}{a_{\mathrm{v}}^\dag}
\newcommand{\acdag}{a_{\mathrm{c}}^\dag}
\newcommand{\avidag}{a_{\mathrm{v,i}}^\dag}
\newcommand{\acidag}{a_{\mathrm{c,i}}^\dag}
\newcommand{\av}{a_{\mathrm{v}}^{\phantom \dag}}
\newcommand{\ac}{a_{\mathrm{c}}^{\phantom \dag}}
\newcommand{\avi}{a_{\mathrm{v,i}}^{\phantom \dag}}
\newcommand{\aci}{a_{\mathrm{c,i}}^{\phantom \dag}}
\newcommand{\cdag}{c^\dag}
\newcommand{\cdagqs}{c^\dag_{\mathrm{q'}}}
\newcommand{\can}{c}
\newcommand{\canqs}{c_{\mathrm{q'}}^{\phantom \dag}}
\newcommand{\ddagq}{d_{\mathrm{q}}^\dag}
\newcommand{\dq}{d_{\mathrm{q}}^{\phantom \dag}}
\newcommand{\dqs}{d_{\mathrm{q'}}^{\phantom \dag}}
\newcommand{\fc}{f^c}
\newcommand{\fv}{f^v}
\newcommand{\fcv}{f^{c/v}}
\newcommand{\plcdagc}{\erw{\cdag\can}}
\newcommand{\plddagqc}{\erw{\ddagq\can}}
\newcommand{\plavaccdag}{\erw{\avdag\ac\cdag}}
\newcommand{\plavacddagq}{\erw{\avdag\ac\ddagq}}
\newcommand{\cdagc}{n_{\mathrm{ph}}}
\newcommand{\ddagqc}{n_{\mathrm{q},0}}
\newcommand{\ddagqsc}{n_{\mathrm{q'},0}}
\newcommand{\avaccdag}{p_1}
\newcommand{\pclass}{p_{\textrm{cl}}}
\newcommand{\avacddagq}{p_{1,\mathrm{q}}}
\newcommand{\plcddagqcdagcdqs}{\erwc{\ddagq\cdag\can\dqs}}
\newcommand{\plddagqdqs}{\erw{\ddagq\dqs}}
\newcommand{\ddagqdqs}{n_{\mathrm{q,q'}}}
\newcommand{\taumeinsspg}{\tau_{\mathrm{sp}_{\mathrm{g}}}^{-1}}
\newcommand{\taumeinsspe}{\tau_{\mathrm{sp}_{\mathrm{e}}}^{-1}}

\begin{document}

\title{Feedback-Induced Steady-State Light Bunching Above the Lasing Threshold}

\author{Franz Schulze}
\email[]{schulze@itp.tu-berlin.de}
\affiliation{Institut f\"ur Theoretische Physik, Nichtlineare Optik und Quantenelektronik, Technische Universit\"at Berlin, Hardenbergstra\ss e 36, 10623 Berlin, Germany}
\author{Benjamin Lingnau}
\affiliation{Institut f\"ur Theoretische Physik, Nichtlineare Dynamik und Kontrolle, Technische Universit\"at Berlin, Hardenbergstra\ss e 36, 10623 Berlin, Germany}
\author{Sven Moritz Hein}
\author{Alexander Carmele}
\affiliation{Institut f\"ur Theoretische Physik, Nichtlineare Optik und Quantenelektronik, Technische Universit\"at Berlin, Hardenbergstra\ss e 36, 10623 Berlin, Germany}
\author{Eckehard Sch\"{o}ll}
\author{Kathy L\"{u}dge}
\affiliation{Institut f\"ur Theoretische Physik, Nichtlineare Dynamik und Kontrolle, Technische Universit\"at Berlin, Hardenbergstra\ss e 36, 10623 Berlin, Germany}
\author{Andreas Knorr}
\affiliation{Institut f\"ur Theoretische Physik, Nichtlineare Optik und Quantenelektronik, Technische Universit\"at Berlin, Hardenbergstra\ss e 36, 10623 Berlin, Germany}

\date{\today}

\begin{abstract}
We develop a full quantum-optical approach for optical self-feedback of a microcavity laser. These miniaturized devices work in a regime between the quantum and classical limit and are test-beds for the differences between a quantized theory of optical self-feedback and the corresponding semiclassical theory.
The light intensity and photon statistics are investigated with and without an external feedback:
We show that in the low-gain limit, where relaxation oscillations do not appear, the recently observed photon bunching in a quantum dot microcavity laser with optical feedback can be accounted for only by the fully quantized model.
By providing a description of laser devices with feedback in the quantum limit we reveal novel insights into the origin of bunching in quantized and semiclassical models.
\end{abstract}


\maketitle

\emph{Introduction}---
Lasers are a cornerstone of modern technology.
They also constitute ideal systems to study a variety of non-linear effects which open possible routes to the exploitation of complex dynamics in applications as well as in fundamental research. Especially semiconductor lasers with external optical feedback can exhibit rich dynamics which depends strongly on the feedback strength or phase and is under intense investigation \cite{Lang1980,Heil2003,Otto2010,Tartwijk1995,Levine1995,Globisch2012}.
Most experimental and theoretical investigations of feedback have focussed on semiconductor lasers involving high numbers of active emitters and photons with output powers in the mW (high gain) regime.
In this regime, a semiclassical treatment of the light field, as in the well established Lang-Kobayashi-model \cite{Lang1980}, is capable of reproducing
rich dynamics observed experimentally \cite{Soriano2013}.
Recent experiments have also considered miniaturized systems such as microcavity lasers which allow for exploring the regime of much lower output intensity and gain, where only a few dozens emitters are involved, and have observed a modified influence of optical self-feedback on the laser statistics \cite{Albert2011}.
In general, feedback in semiconductor lasers typically induces chaotic emission resulting in a bunched photon statistics, i.e., a photon-photon correlation $\gtwo>1$ compared to the pure lasing limit $\gtwo=1$.
Such classical radiation with $\gtwo\geq1$ can usually be described by semiclassical models using quantum theory for the emitters but treating the field classically.
On the other hand, a low intensity/low gain situation, where only a small number of emitters are involved, typically requires a full quantum description \cite{Loudon2003}.
Therefore, the range of validity of the semiclassical description is not clear.
\\
In this Rapid Communication, we develop a fully quantized theory of optical self-feedback in a low-gain regime
characteristic of a microcavity laser operating between the quantum and the semiclassical limit.
We compare it to a semiclassical approach and find qualitative differences in the light field statistics above the lasing threshold in the low-gain regime.
Here, we define the low-gain regime by the absence of turn-on laser relaxation oscillations: In this regime, the gain is not sufficient to boost the intensity at switch-on above its stationary value before the cavity losses start inhibiting its growth.
As shown in this work, the absence of undamped oscillations, i.e., continuous wave emission under feedback, results in the following behavior:
While the semiclassical theory shows no bunching behavior of the light statistics above the lasing threshold under optical self-feedback, the quantum-optical approach reveals such bunching also for stationary intensity in the low-gain regime.
We identify two different origins of feedback-induced light bunching above the lasing threshold:
One is connected to nonstationary behavior of the mean light field intensity and strongly connected to a random phase superposition of classical coherent waves.
The other stems from the influence of feedback on quantum-mechanical photon correlations and leads to bunching of the light field statistics even if in a steady state.
To connect our calculation to recent experimental results, we address a semiconductor quantum dot microcavity laser, however, our results apply to all lasers with feedback.

This Rapid Communication is structured as follows:
First, we introduce a fully quantized model of a semiconductor microcavity laser with optical self-feedback.
For comparison, the corresponding semiclassical model, based on the Lang-Kobayashi equations is also introduced and the connection between the classical and quantum-mechanical approach is discussed.
Second, both models are compared in a low-gain regime where both approaches meet at the border of the transition between classical and quantum behavior of a quantum dot laser: in this limit, qualitative differences in the photon statistics are investigated.
In contrast to the common assumption that a randomly phased superposition of coherent waves leading to bunched classical light is established in the semiclassical description, no such behavior is observed in the low-gain regime of a laser and only the quantized description reproduces the measured effect of bunching above the lasing threshold.
We conclude by outlining the origin of this different behavior.

\begin{figure}[htb]
\includegraphics[width=2.4in,clip]{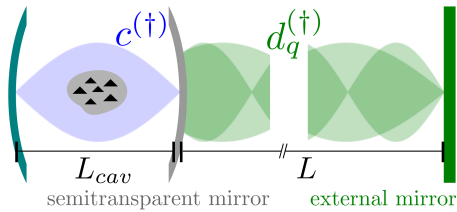}
\caption{\label{fig_sketch}(Color online) Model system of $\NQD$ quantum dots inside an optical microcavity of length $L_{\textrm{cav}}$ with an external optical mirror at a distance $L\gg L_{\textrm{cav}}$.}
\end{figure}

\emph{Model System}---
We consider a quantum dot laser model \cite{Gies2007,Otto2010,Leymann2013} with a single lasing mode and extend it by a dynamical description of the light-light interaction of the cavity and the external light field (see Fig.~\ref{fig_sketch}).
This extension will enable the treatment of an external mirror and its influence on the dynamical and statistical properties of the strong lasing mode of a possible experimental setup \cite{Albert2011}.
The laser transition of $\NQD$ quantum dots is assumed to be in resonance with the quantized microcavity mode ($c^{(\dag)}$). This mode is coupled to a continuum of external modes ($d_q^{(\dag)}$) which are influenced by the presence of a dielectric mirror at distance $L$.
The full system Hamiltonian reads
$H	= H_0 + H_{\mathrm{EL}} + H_{\mathrm{LL}}\mathrm{\, ,}$
where $H_0$ includes the free dynamics of the electrons and photons inside and outside the laser cavity, see Fig.~\ref{fig_sketch}:
\begin{align}\label{sys_H_0}
H_0	&= \sumi(\epsilon_{\mathrm{v,i}} \avidag\avi + \epsilon_{\mathrm{c,i}} \acidag\aci)
\\	&+ \hbar \sumqs \omnull \cdagqs\canqs + \hbar \sumq \omq \ddagq\dq\,\,.
\nonumber
\end{align}
$\avi$ ($\aci$) and $\avidag$ ($\acidag$) are the annihilation and creation operators of electrons in the valence (conduction) band of the ith quantum dot, respectively. $\canqs$ ($\dq$) and $\cdagqs$ ($\ddagq$) annihilate and create, respectively a cavity (external) photon with momentum $q'$ ($q$) and frequency $\omnull$ ($\omega_q$).
In the final equations, interactions between the electronic system and non-lasing cavity modes with $q' \neq q_0$ ($q_0$ denoting the dominant lasing mode) will be summarized in the spontaneous emission factor $\beta$.
It describes the fraction of spontaneous emission into the lasing mode relative to the total spontaneous emission and will be treated as a parameter \cite{Gies2007}.
The momentum index $q_0$ of the dominant cavity mode will be omitted in the following for the sake of clarity.
For simplicity, the energy gap between the conduction and valence band ground levels for $\NQD$ quantum dots is assumed to be in resonance with the energy of the lowest cavity mode, i.e., $\epsilon_{\mathrm{c,i}} - \epsilon_{\mathrm{v,i}} = \hbar\omnull$.
The light-matter interaction between the resonant QD electronic levels and the cavity is treated within the dipole approximation and reads in the rotating wave approximation:
\begin{equation}\label{sys_HQO}
H_{\mathrm{EL}}	= - \hbar \sumi \Mi \avidag\aci\cdag + \text{h. c. ,}
\end{equation}
with the coupling element $\Mi = - \Msi$ \cite{Armen2006}.
A number of $\NQD$ equal quantum dots will be assumed in the detailed calculations and the quantum dot index $i$ will therefore be neglected.
The self-feedback of the cavity modes is introduced by including the interaction $H_{\mathrm{LL}}$ between the cavity and external light field: 
\begin{equation}\label{sys_Hll}
H_{\mathrm{LL}} = - \hbar \sumq \Gsq \ddagq \can + \text{h. c. ,}
\end{equation}
with the coupling element $\Gsq = - \Gq$ \cite{Dorner2002} which is, in general, momentum dependent (see discussion below).
Again, we neglect non-energy conserving terms by applying the rotating wave approximation.

Next, we derive the feedback controlled laser dynamics.
The equation of motion approach is employed to derive dynamical equations for expectation values of observables such as cavity photon number and photon-photon correlations:
$\dt \erw{A} = \frac{\I}{\hbar} \erw{\big[H,A\big]_-}$.
The hierarchy problem \cite{Trimborn2009,Kira2006} emerges due to our model system with feedback above the one-photon limit \cite{Carmele2013}.
Here, it is treated in the correlation expansion approach \cite{Wang1985,Axt1998,Kira2006,Gies2007} which is valid for photon numbers well above unity.
We will take into account correlations up to the fourth order in the light coupling element $M$ which is mandatory for an investigation of the second order auto-correlation function $g^{(2)}(0) = \erw{\cdag\cdag\can\can}/\erw{\cdag\can}^2$.
While the constituents of the Hamiltonian given by Eqs.~(\ref{sys_H_0}) and (\ref{sys_HQO}) correspond to a QD laser theory \cite{Gies2007}, the light-light coupling Hamiltonian [Eq.~(\ref{sys_Hll})] includes self-feedback by the external mirror on a quantum-optical level.
We now discuss the two levels of description (fully quantized and semiclassical model).

\emph{Fully Quantized Model}---
The coupling between the cavity modes and the external modes is taken in the limit of a good cavity. This allows for the separate treatment of cavity and external modes in contrast to the leaky cavity case \cite{Barnett1988}.  
In this approximation we assume the mode structure of the laser cavity and the external modes to be independent.
Then, the coupling between both fields is described by a momentum dependent coupling element $\Gq$, cf. Eq.~(\ref{sys_Hll}), which carries the momentum dependencies of the cavity and external light field at the semi-transparent cavity mirror (see Fig.~\ref{fig_sketch}).
The matrix element $\Gq$ is taken as constant for coupling into free space \cite{Dorner2002} at optical frequencies, i.e., $\Gq = G_0$.
In contrast, in the case of an external mirror a momentum dependent coupling element with a sinusoidal dependence on the spatial phase $q L$ resembles a mirror at a distance~$L$~\cite{Dorner2002}, i.e. $\Gq = G_0 \sqrt{2} \sin{\big(q L\big)}$.

For the derivation of the light intensity and the photon statistics, the interaction between the cavity and the external light field is taken into account up to the second order in the coupling element $G_q$, i.e., expectation values of up to two external photon operators ($d^{(\dag)}_q$). This treatment includes the physically relevant photon densities \emph{and} photon coherences of the external light field and enables a numerical solution of the dynamical equations.
Note that expectation values which contain two external photon operators and at least one more cavity photon operator, for example $\plcddagqcdagcdqs$, are not damped by the light-light interaction $H_{\textrm{LL}}$ due to our truncation procedure. A phenomenological Markovian loss rate $\kappah$ is therefore introduced to simulate higher order correlations.
Its value is chosen to resemble the losses introduced by the cavity coupling to the external field and does not qualitatively influence the dynamics over a wide range of situations discussed here.
Losses for the external field are introduced by a loss rate $\kappaext$ which is included into the dynamical equations by a Lindblad approach.
Starting from the cavity photon density $\plcdagc=\cdagc$ and the photon-photon correlation function $\gtwo = \erw{\cdag\cdag\can\can}/\erw{\cdag\can}^2$ both couple to a nonlinear hierarchy of equations of motion.
Our quantized description of optical feedback foots on a basic set of equations \cite{Carmele2013} which is expanded to higher correlations for the case of large photon numbers \cite{Gies2007} to describe the laser action.
The full set of equations is given in the supplementary material.
Here, we discuss the underlying system of equations [see Eqs.~(\ref{eqs_cdagc_text})-(\ref{eqs_ddagqdqs_text})].
The cavity photon density $\cdagc$ couples to the total number of $\NQD$ quantum dots and to the photon-assisted polarization $\avaccdag= \plavaccdag$ and to the photon transfer amplitudes $\ddagqc = \plddagqc$:
\begin{align}
\label{eqs_cdagc_text}
\dt \cdagc			= &- 2 \cimag{\M \avaccdag 
					} \NQD
 				  + 2 \cimag{\sumq \Gsq \ddagqc }
\end{align}
In turn, the photon-assisted polarization $\avaccdag$ couples also to the external photon-assisted polarizations $\avacddagq = \plavacddagq$ [see Eqs.~(\ref{eqs_avaccdag_text})-(\ref{eqs_avacddagq_text})]
which are both driven by the electronic occupations of conduction/valence band $f^{c/v} = \erw{a_{c/v}^\dag a^{\phantom\dag}_{c/v}}$ and damped by pure dephasing $\gammaPD$: 
\begin{align}
\label{eqs_avaccdag_text}
(\dt &+ \gammaPD 
) \avaccdag	= 
				  - \I \sumq \Gsq \avacddagq
\\
&				  - \I \Ms 
				  [
					\fc (1-\fv) 
				  +
					(\fc- \fv)  \cdagc
				  +
				  \erw{\phantom .}|_{\textrm{corr}}
				  ]
				  \nonumber
\end{align}
\begin{align}
\label{eqs_avacddagq_text}
   (\dt &+ \GAMMAextPD + \I\DELTAcvq) \avacddagq		=   
\\
				    &- \I \Ms 
				    [
					(\fc-\fv) \ddagqc
				    +
				  \erw{\phantom .}|_{\textrm{corr}}
				    ]
					\nonumber
  				    - \I \Gq 
					\avaccdag 
					\nonumber
\end{align}
where $\Delta^{m}_{n} = (\omega_{m} - \omega_{n})$ is a detuning and $\erw{\phantom .}|_{\textrm{corr}}$ abbreviates the coupling to higher order terms which are the connections to the full set of equations.
The expectation value of the photon transfer amplitude $\ddagqc$ depends on the state of the cavity light field $\cdagc$ and the external light field including all photon densities and external mode coherences $\ddagqdqs = \plddagqdqs$:
\begin{align} 
\label{eqs_ddagqc_text}
    (\dt &+ \kappaext - \I\DELTAomqnull) \ddagqc			=
\\
    					&+ \I \M 
					\avacddagq  \NQD
    				    - \I \Gq 
					\cdagc 
				    + \I \sumqs \Gqs \ddagqdqs\,\,,
					\nonumber
\end{align}
\begin{align}
\label{eqs_ddagqdqs_text}
(\dt + 2 \kappaext - \I\DELTAomqomqs) \ddagqdqs		=   	    &+ \I  \Gsqs 
					\ddagqc 
 				    - \I  \Gq 
					\ddagqsc^* \,\,.
\end{align}

\emph{Semiclassical Model}---
The semiclassical model is derived by factorizing the expectation values of the cavity intensity $\erw{\cdag\can}$ and photon-assisted polarization $\erw{\acdag\av\can}$ and neglecting quantum-mechanical correlations, i.e., setting $\erw{\cdag\can} = \big|\erw{\can}\big|^2$ and $\erw{\acdag\av\can} = \erw{\acdag\av}\erw{\can} = \pclass \erw{\can}$.
Initial driving of the coherent fields, e.g. $\erw{\can}$, is induced by modelling spontaneous emission in a classical framework~[20]: We use
complex Gaussian white noise $\xi(t)$ with $\erwt{\xi(t)\xi(t')} = \delta(t-t')$ and $\erwt{\creal{\xi(t)}\cimag{\xi(t')}} = 0$, where $\erwt{\phantom .}$ is the temporal average.
\textit{It is the main point of our work that this classical noise, usually successful to describe photon bunching in lasers, fails for feedback in the low-gain regime.}
The time-delay $\tau$ introduced by the external cavity results in a dynamical equation for the classical cavity field $\erw{\can}$ which is of the Lang-Kobayashi type:
\begin{align}
\dt \erw{\can} =
&+ \kappa S e^{-\I \phi} \erw{\can}(t-\tau) - \kappa \erw{\can}
\\
&+ \I \NQD \M \pclass + \sqrt{\beta \NQD \fc (1 - \fv) \taumeinsspg} \xi(t)
\nonumber
\end{align}
\begin{align}
\dt \pclass =
&- \I \M (\fc - \fv) \erw{\can} - \gammaPD \pclass
\end{align}
\begin{align}
\dt \fcv =
&\pm 2 \M \cimag{\pclass \erw{\can}} \mp \fc(1-\fv)\taumeinsspg \pm \erw{\phantom .}|_{\textrm{pump}}
\end{align}
where the rates for cavity loss rate $\kappa$ and spontaneous emission of the ground state $\taumeinsspg$ are included.
Here, the delay time $\tau = 2 L/c$ and the feedback phase $\phi = \tau \omnull$ of the semiclassical model are directly related to the feedback length $L$ in the quantized model. The feedback strength $S$ in the semiclassical description is connected to the losses in the quantized external field $\kappaext$ by $S = e^{-\kappaext \tau}$.
The \gtwofunction\,\,of the semiclassical model is calculated as $\gtwo = \erwt{|\erw{\can}|^4}/\erwt{|\erw{\can}|^2}^2$ \cite{Loudon2003}.

\emph{Comparison of the fully quantized and the semiclassical model}---
Next, using numerical solutions of the dynamical equations, we discuss the similarities and differences of both models.
The correlation expansion used is an improvement of the mean field approach (which is strictly valid only in the semiclassical range) by including a certain level of N-particle correlations.
Thus, for too few quantum dots, the correlation expansion breaks down since correlations are too dominating.
The actual number of quantum dots that can be treated depends on the parameter range and is benchmarked by an independent two level calculation.
To be specific, the quantized model used in this work (with a truncation on the four-particle level, and operating in the investigated good cavity limit) exhibits coherent emission in accordance with standard laser theory down to $\geq 40$ emitters.
Higher order truncations or different truncation schemes~[20] can decrease the QD number further but are not numerically feasible.
Similarly, because of the lower order truncation 
the semiclassical model is only capable of describing a laser with $\geq 2000$ emitters, otherwise no sufficient gain can be achieved.
Therefore, we compare both approaches considering $2000$ QDs.
On the other hand, we clearly have to distinguish between the theoretical comparison of the two models and an experiment \cite{Albert2011} done in the limit of approximately 10 emitters.
However, since we can use the quantum model down to few tens of emitters obtaining similar results, this gives strong evidence how to understand the experiment in Ref.~\cite{Albert2011}.

\begin{figure}[htb]
\includegraphics[width=3.4in,clip]{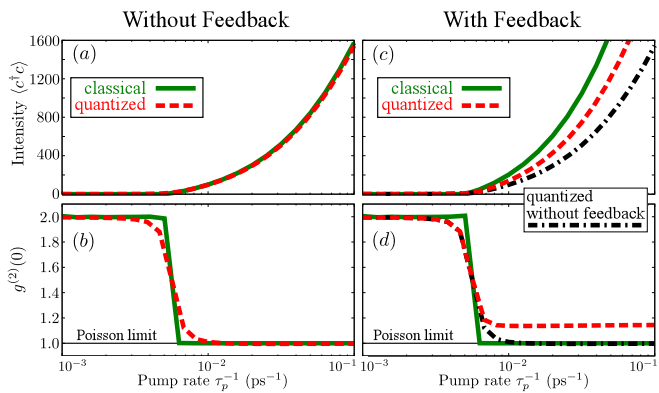}
\caption{(Color online) Calculated input-output curves of the quantum dot laser are shown (a), (b) without and (c), (d) with external feedback. No significant deviations occur between the semiclassical (green solid) and fully quantized model (red dashed) without external feedback (a), (b). With external feedback (c), (d), the mean intensities (c) are enhanced with respect to the case without feedback (black dash-dotted) and the quantum-optical calculation exhibits super-Poissonian statistics above the lasing threshold (d).
\label{fig_comparison_cl_QO}}
\end{figure}

Note that these parameters (see Table~\ref{table_parameters}) still constitute the low-gain regime and lie well below high-gain laser QD numbers of hundred thousands \cite{Otto2010}.
Figure~\ref{fig_comparison_cl_QO} shows the laser input-output curves without (a,b) and with (c,d) external feedback for the same set of parameters.
All figures display the numerical solutions of the semiclassical (green solid) and the quantum-optical (red dashed) model.
Agreement is visible in the intensity output without external feedback (see Fig.~\ref{fig_comparison_cl_QO}(a)).
Also the \gtwofunction\, is in good agreement (see Fig.~\ref{fig_comparison_cl_QO}(b)).
Differences in the steepness of the drop at the lasing threshold from thermal to Poissonian statistics arise due to the different treatment of spontaneous emission in both models.
The application of an external optical feedback leads in both models to a rise in the output intensities (Fig.~\ref{fig_comparison_cl_QO}(c)) with respect to the case without external mirror (black dash-dotted).
This can be understood in the following way: The external feedback lowers the overall optical loss rate of the laser since a certain amount of coupled-out light is fed back into the laser. This leads to a higher mean intensity.
However, and this is the main result of this work, we find a clear \textit{qualitative} difference between the quantized and semiclassical description in the field statistics above the lasing threshold if optical self-feedback is applied (see Fig.~\ref{fig_comparison_cl_QO}(d)).  The quantized description exhibits a rise in the autocorrelation function from Poissonian photon statistics, i.e. $\gtwo = 1$, to super-Poissonian statistics, i.e. $\gtwo > 1$.
In contrast, the semiclassical description predicts a fully coherent Poissonian light emission.
We will now investigate the reason why this difference between the two approaches exists:
In a semiclassical description, photon bunching occurs only for nonstationary intensities.
Since the mean intensity is stationary, the quantum-optical result of bunching is in strong contrast to a classical description \cite{Loudon2003}.
We have checked that bunching of the photon statistics ($\gtwo>1$) above the lasing threshold can also be achieved in the semiclassical theory but it requires a nonstationary time trace of the mean light field intensity.
Such nonstationary behavior has been observed and is well described by semiclassical theories of mW-lasers; it relies on the undamping of relaxation oscillations via a Hopf bifurcation \cite{Mork1992}, which can subsequently lead to the appearance of chaotic dynamics.
In the regime between the extreme quantum limit and classical conditions, treated in this work, the gain medium does not provide the high gain necessary for the appearance of a Hopf bifurcation.
Therefore, nonstationary behavior of the coherent light field cannot arise and consequently no super-Poissonian statistics of the light field are observed in the semiclassical model.
In contrast, the quantized description also in the low-gain limit exhibits clear bunching of the light field above the lasing threshold which is an experimentally measurable feature~\cite{Albert2011}.
A detailed analysis of the quantum optical equations in the limit of weak feedback shows that the photon density and the photon correlations underly different feedback contributions.
Photon-photon correlations are more sensitive to feedback compared to photon densities:
This results in an imbalance between both quantities compared to the pure laser emission and thus a small bunching above the lasing threshold in a steady state.
One can say that feedback modifies the exact Glauber state of the running laser to be more chaotic/bunched.

\begin{figure}[htb]
\includegraphics[width=3.4in,clip]{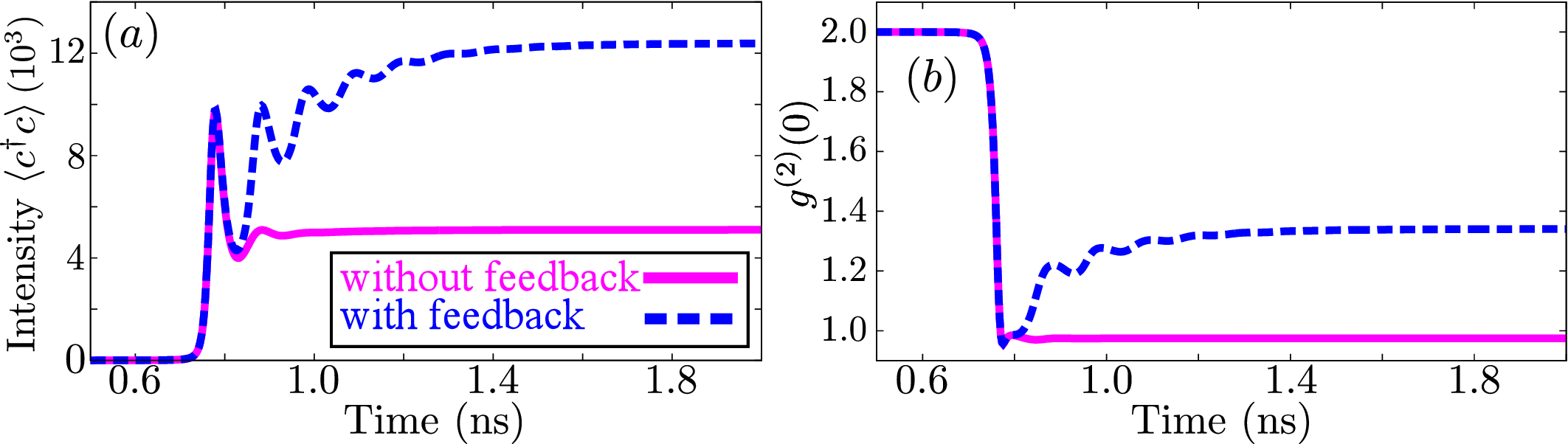}
\caption{(Color online) Quantized approach: Time transients of (a) the intensity and (b) \gtwofunction\,\,without (purple solid) and with (blue dashed) feedback.
(a) Relaxation oscillations are visible in the intensity and are prolonged by optical self-feedback.
(b) Super-Poissonian statistics are approached by the \gtwofunction\,\,in the stationary limit. Parameters: $\NQD = 1.8\times10^{6}, \beta = 1.\times10^{-3}, S=0.65$
\label{fig_ueber_zeit}}
\end{figure}

A fully quantized treatment in the classical parameter regime exhibiting nonstationary intensities including classical chaos is not feasible because of the enormous computational requirements.
Nevertheless, a reduction of the relaxation oscillation damping with increasing optical feedback can be described also by the quantum-optical approach for parameters of a high-gain medium (see Fig.~\ref{fig_ueber_zeit}).
Figure~\ref{fig_ueber_zeit}(b) shows clearly that the unsteady behavior of the light field intensity influences the \gtwofunction\,\,\emph{and} a super-Poissonian statistics is reached even for a steady state intensity.

\emph{Conclusion}---
The quantized approach to optical self-feedback offers a novel point of view on the intriguing relation between semiclassical complex dynamics and the quantum-mechanical description of photon statistics.
Referring to these observations, we conclude that feedback-induced light bunching phenomena above the lasing threshold can have two different origins: (i) The chaotic or oscillatory dynamics of the mean light field intensity arising from feedback, described in semiclassical models. (ii) The influence of feedback
as a coupling mechanism between the photon density and photon-photon correlation
which is only described by the fully quantized approach and results in a bunching of the light field statistics above the lasing threshold in a steady state.
This behavior is in full agreement with a recent experiment \cite{Albert2011} on few emitter, low-gain quantum dot lasers.
\section*{Acknowledgements}
The authors acknowledge support from Deutsche Forschungsgemeinschaft through SFB 910 (project B1) and SFB 787 (project B2).
We wish to thank S. Reitzenstein, I. Kanter, J. Kabuss, C. Hopfmann, and F. Gericke for helpful discussions.

\begin{table}[htb]
\begin{center}
\begin{tabular}{cc|cc}
\hline
\hline
$\beta$ 			& $1 \times 10^{-4}$	& $\kappa^{-1},\kappah^{-1}$	& $22\,\text{ps}$	\\
$\NQD$ 				& $2000$		& $\taumeinsspe$, $\taumeinsspg$& $\frac{1}{50000}\,\text{fs}^{-1}$\\
$\tau$				& $90\,\text{ps}$	& $\gammaPD$			& $\frac{1.36\,\text{meV}}{\hbar}$\\
$S = e^{- \kappaext \tau}$ 	& $0.5$			& 				& \\
\hline
\hline
\end{tabular}
\end{center}
\caption{Simulation parameters\label{table_parameters}}
\end{table}


\begin{thebibliography}{20}
\expandafter\ifx\csname natexlab\endcsname\relax\def\natexlab#1{#1}\fi
\expandafter\ifx\csname bibnamefont\endcsname\relax
  \def\bibnamefont#1{#1}\fi
\expandafter\ifx\csname bibfnamefont\endcsname\relax
  \def\bibfnamefont#1{#1}\fi
\expandafter\ifx\csname citenamefont\endcsname\relax
  \def\citenamefont#1{#1}\fi
\expandafter\ifx\csname url\endcsname\relax
  \def\url#1{\texttt{#1}}\fi
\expandafter\ifx\csname urlprefix\endcsname\relax\def\urlprefix{URL }\fi
\providecommand{\bibinfo}[2]{#2}
\providecommand{\eprint}[2][]{\url{#2}}

\bibitem[{\citenamefont{Lang and Kobayashi}(1980)}]{Lang1980}
\bibinfo{author}{\bibfnamefont{R.}~\bibnamefont{Lang}} \bibnamefont{and}
  \bibinfo{author}{\bibfnamefont{K.}~\bibnamefont{Kobayashi}},
  \bibinfo{journal}{IEEE J. Quantum Electron.} \textbf{\bibinfo{volume}{16}},
  \bibinfo{pages}{347} (\bibinfo{year}{1980}).

\bibitem[{\citenamefont{Heil et~al.}(2003)\citenamefont{Heil, Fischer,
  Els{\"{a}\ss er}, Krauskopf, Green, and Gavrielides}}]{Heil2003}
\bibinfo{author}{\bibfnamefont{T.}~\bibnamefont{Heil}},
  \bibinfo{author}{\bibfnamefont{I.}~\bibnamefont{Fischer}},
  \bibinfo{author}{\bibfnamefont{W.}~\bibnamefont{Els{\"{a}\ss er}}},
  \bibinfo{author}{\bibfnamefont{B.}~\bibnamefont{Krauskopf}},
  \bibinfo{author}{\bibfnamefont{K.}~\bibnamefont{Green}}, \bibnamefont{and}
  \bibinfo{author}{\bibfnamefont{A.}~\bibnamefont{Gavrielides}},
  \bibinfo{journal}{Phys. Rev. E} \textbf{\bibinfo{volume}{67}},
  \bibinfo{pages}{066214} (\bibinfo{year}{2003}).

\bibitem[{\citenamefont{Otto et~al.}(2010)\citenamefont{Otto, L\"{u}dge, and
  Sch\"{o}ll}}]{Otto2010}
\bibinfo{author}{\bibfnamefont{C.}~\bibnamefont{Otto}},
  \bibinfo{author}{\bibfnamefont{K.}~\bibnamefont{L\"{u}dge}},
  \bibnamefont{and}
  \bibinfo{author}{\bibfnamefont{E.}~\bibnamefont{Sch\"{o}ll}},
  \bibinfo{journal}{Phys. Stat. Sol. b} \textbf{\bibinfo{volume}{247}},
  \bibinfo{pages}{829} (\bibinfo{year}{2010}).

\bibitem[{\citenamefont{van Tartwijk and Lenstra}(1995)}]{Tartwijk1995}
\bibinfo{author}{\bibfnamefont{G.~H.~M.} \bibnamefont{van Tartwijk}}
  \bibnamefont{and} \bibinfo{author}{\bibfnamefont{D.}~\bibnamefont{Lenstra}},
  \bibinfo{journal}{J. Quantum Semiclass. Opt.} \textbf{\bibinfo{volume}{7}},
  \bibinfo{pages}{87} (\bibinfo{year}{1995}).

\bibitem[{\citenamefont{Levine et~al.}(1995)\citenamefont{Levine, van Tartwijk,
  Lenstra, and Erneux}}]{Levine1995}
\bibinfo{author}{\bibfnamefont{A.~M.} \bibnamefont{Levine}},
  \bibinfo{author}{\bibfnamefont{G.~H.~M.} \bibnamefont{van Tartwijk}},
  \bibinfo{author}{\bibfnamefont{D.}~\bibnamefont{Lenstra}}, \bibnamefont{and}
  \bibinfo{author}{\bibfnamefont{T.}~\bibnamefont{Erneux}},
  \bibinfo{journal}{Phys. Rev. A} \textbf{\bibinfo{volume}{52}},
  \bibinfo{pages}{R3436} (\bibinfo{year}{1995}).

\bibitem[{\citenamefont{Globisch et~al.}(2012)\citenamefont{Globisch, Otto,
  L\"{u}dge, and Sch\"{o}ll}}]{Globisch2012}
\bibinfo{author}{\bibfnamefont{B.}~\bibnamefont{Globisch}},
  \bibinfo{author}{\bibfnamefont{C.}~\bibnamefont{Otto}},
  \bibinfo{author}{\bibfnamefont{K.}~\bibnamefont{L\"{u}dge}},
  \bibnamefont{and}
  \bibinfo{author}{\bibfnamefont{E.}~\bibnamefont{Sch\"{o}ll}},
  \bibinfo{journal}{Phys. Rev. E} \textbf{\bibinfo{volume}{86}},
  \bibinfo{pages}{046201} (\bibinfo{year}{2012}).

\bibitem[{\citenamefont{Soriano et~al.}(2013)\citenamefont{Soriano,
  Garcia-Ojalvo, Mirasso, and Fischer}}]{Soriano2013}
\bibinfo{author}{\bibfnamefont{M.~C.} \bibnamefont{Soriano}},
  \bibinfo{author}{\bibfnamefont{J.}~\bibnamefont{Garcia-Ojalvo}},
  \bibinfo{author}{\bibfnamefont{C.~R.} \bibnamefont{Mirasso}},
  \bibnamefont{and} \bibinfo{author}{\bibfnamefont{I.}~\bibnamefont{Fischer}},
  \bibinfo{journal}{Rev. Mod. Phys.} \textbf{\bibinfo{volume}{85}},
  \bibinfo{pages}{421} (\bibinfo{year}{2013}).

\bibitem[{\citenamefont{Albert et~al.}(2011)\citenamefont{Albert, Hopfmann,
  Reitzenstein, Schneider, H\"{o}fling, Worschech, Kamp, Kinzel, Forchel, and
  Kanter}}]{Albert2011}
\bibinfo{author}{\bibfnamefont{F.}~\bibnamefont{Albert}},
  \bibinfo{author}{\bibfnamefont{C.}~\bibnamefont{Hopfmann}},
  \bibinfo{author}{\bibfnamefont{S.}~\bibnamefont{Reitzenstein}},
  \bibinfo{author}{\bibfnamefont{C.}~\bibnamefont{Schneider}},
  \bibinfo{author}{\bibfnamefont{S.}~\bibnamefont{H\"{o}fling}},
  \bibinfo{author}{\bibfnamefont{L.}~\bibnamefont{Worschech}},
  \bibinfo{author}{\bibfnamefont{M.}~\bibnamefont{Kamp}},
  \bibinfo{author}{\bibfnamefont{W.}~\bibnamefont{Kinzel}},
  \bibinfo{author}{\bibfnamefont{A.}~\bibnamefont{Forchel}}, \bibnamefont{and}
  \bibinfo{author}{\bibfnamefont{I.}~\bibnamefont{Kanter}},
  \bibinfo{journal}{Nat. Commun.} \textbf{\bibinfo{volume}{2}},
  \bibinfo{pages}{366} (\bibinfo{year}{2011}).

\bibitem[{\citenamefont{Loudon}(2003)}]{Loudon2003}
\bibinfo{author}{\bibfnamefont{R.}~\bibnamefont{Loudon}},
  \emph{\bibinfo{title}{The Quantum Theory of Light}}
  (\bibinfo{publisher}{Oxford: Oxford Science Publications},
  \bibinfo{year}{2003}), \bibinfo{edition}{3rd} ed.

\bibitem[{\citenamefont{Gies et~al.}(2007)\citenamefont{Gies, Wiersig, Lorke,
  and Jahnke}}]{Gies2007}
\bibinfo{author}{\bibfnamefont{C.}~\bibnamefont{Gies}},
  \bibinfo{author}{\bibfnamefont{J.}~\bibnamefont{Wiersig}},
  \bibinfo{author}{\bibfnamefont{M.}~\bibnamefont{Lorke}}, \bibnamefont{and}
  \bibinfo{author}{\bibfnamefont{F.}~\bibnamefont{Jahnke}},
  \bibinfo{journal}{Phys. Rev. A} \textbf{\bibinfo{volume}{75}},
  \bibinfo{pages}{013803} (\bibinfo{year}{2007}).

\bibitem[{\citenamefont{Leymann et~al.}(2013)\citenamefont{Leymann, Hopfmann,
  Albert, Foerster, Khanbekyan, Schneider, H{\"o}fling, Forchel, Kamp, Wiersig
  et~al.}}]{Leymann2013}
\bibinfo{author}{\bibfnamefont{H.~A.~M.} \bibnamefont{Leymann}},
  \bibinfo{author}{\bibfnamefont{C.}~\bibnamefont{Hopfmann}},
  \bibinfo{author}{\bibfnamefont{F.}~\bibnamefont{Albert}},
  \bibinfo{author}{\bibfnamefont{A.}~\bibnamefont{Foerster}},
  \bibinfo{author}{\bibfnamefont{M.}~\bibnamefont{Khanbekyan}},
  \bibinfo{author}{\bibfnamefont{C.}~\bibnamefont{Schneider}},
  \bibinfo{author}{\bibfnamefont{S.}~\bibnamefont{H{\"o}fling}},
  \bibinfo{author}{\bibfnamefont{A.}~\bibnamefont{Forchel}},
  \bibinfo{author}{\bibfnamefont{M.}~\bibnamefont{Kamp}},
  \bibinfo{author}{\bibfnamefont{J.}~\bibnamefont{Wiersig}},
  \bibnamefont{et~al.}, \bibinfo{journal}{Phys. Rev. A}
  \textbf{\bibinfo{volume}{87}}, \bibinfo{pages}{053819}
  (\bibinfo{year}{2013}).

\bibitem[{\citenamefont{Armen and Mabuchi}(2006)}]{Armen2006}
\bibinfo{author}{\bibfnamefont{M.~A.} \bibnamefont{Armen}} \bibnamefont{and}
  \bibinfo{author}{\bibfnamefont{H.}~\bibnamefont{Mabuchi}},
  \bibinfo{journal}{Phys. Rev. A} \textbf{\bibinfo{volume}{73}},
  \bibinfo{pages}{063801} (\bibinfo{year}{2006}).

\bibitem[{\citenamefont{Dorner and Zoller}(2002)}]{Dorner2002}
\bibinfo{author}{\bibfnamefont{U.}~\bibnamefont{Dorner}} \bibnamefont{and}
  \bibinfo{author}{\bibfnamefont{P.}~\bibnamefont{Zoller}},
  \bibinfo{journal}{Phys. Rev. A} \textbf{\bibinfo{volume}{66}},
  \bibinfo{pages}{023816} (\bibinfo{year}{2002}).

\bibitem[{\citenamefont{Trimborn et~al.}(2009)\citenamefont{Trimborn, Witthaut,
  and Korsch}}]{Trimborn2009}
\bibinfo{author}{\bibfnamefont{F.}~\bibnamefont{Trimborn}},
  \bibinfo{author}{\bibfnamefont{D.}~\bibnamefont{Witthaut}}, \bibnamefont{and}
  \bibinfo{author}{\bibfnamefont{H.~J.} \bibnamefont{Korsch}},
  \bibinfo{journal}{Phys. Rev. A} \textbf{\bibinfo{volume}{79}},
  \bibinfo{pages}{013608} (\bibinfo{year}{2009}).

\bibitem[{\citenamefont{Kira and Koch}(2006)}]{Kira2006}
\bibinfo{author}{\bibfnamefont{M.}~\bibnamefont{Kira}} \bibnamefont{and}
  \bibinfo{author}{\bibfnamefont{S.~W.} \bibnamefont{Koch}},
  \bibinfo{journal}{Phys. Rev. A} \textbf{\bibinfo{volume}{73}},
  \bibinfo{pages}{013813} (\bibinfo{year}{2006}).

\bibitem[{\citenamefont{Carmele et~al.}(2013)\citenamefont{Carmele, Kabuss,
  Schulze, Reitzenstein, and Knorr}}]{Carmele2013}
\bibinfo{author}{\bibfnamefont{A.}~\bibnamefont{Carmele}},
  \bibinfo{author}{\bibfnamefont{J.}~\bibnamefont{Kabuss}},
  \bibinfo{author}{\bibfnamefont{F.}~\bibnamefont{Schulze}},
  \bibinfo{author}{\bibfnamefont{S.}~\bibnamefont{Reitzenstein}},
  \bibnamefont{and} \bibinfo{author}{\bibfnamefont{A.}~\bibnamefont{Knorr}},
  \bibinfo{journal}{Phys. Rev. Lett.} \textbf{\bibinfo{volume}{110}},
  \bibinfo{pages}{013601} (\bibinfo{year}{2013}).

\bibitem[{\citenamefont{Wang and Cassing}(1985)}]{Wang1985}
\bibinfo{author}{\bibfnamefont{S.-J.} \bibnamefont{Wang}} \bibnamefont{and}
  \bibinfo{author}{\bibfnamefont{W.}~\bibnamefont{Cassing}},
  \bibinfo{journal}{Ann. Phys.} \textbf{\bibinfo{volume}{159}},
  \bibinfo{pages}{328} (\bibinfo{year}{1985}).

\bibitem[{\citenamefont{Axt and Mukamel}(1998)}]{Axt1998}
\bibinfo{author}{\bibfnamefont{V.~M.} \bibnamefont{Axt}} \bibnamefont{and}
  \bibinfo{author}{\bibfnamefont{S.}~\bibnamefont{Mukamel}},
  \bibinfo{journal}{Rev. Mod. Phys.} \textbf{\bibinfo{volume}{70}},
  \bibinfo{pages}{145} (\bibinfo{year}{1998}).

\bibitem[{\citenamefont{Barnett and Radmore}(1988)}]{Barnett1988}
\bibinfo{author}{\bibfnamefont{S.~M.} \bibnamefont{Barnett}} \bibnamefont{and}
  \bibinfo{author}{\bibfnamefont{P.~M.} \bibnamefont{Radmore}},
  \bibinfo{journal}{Opt. Commun.} \textbf{\bibinfo{volume}{68}},
  \bibinfo{pages}{364} (\bibinfo{year}{1988}).

\bibitem[{\citenamefont{M{\o}rk et~al.}(1992)\citenamefont{M{\o}rk, Tromborg,
  and Mark}}]{Mork1992}
\bibinfo{author}{\bibfnamefont{J.}~\bibnamefont{M{\o}rk}},
  \bibinfo{author}{\bibfnamefont{B.}~\bibnamefont{Tromborg}}, \bibnamefont{and}
  \bibinfo{author}{\bibfnamefont{J.}~\bibnamefont{Mark}},
  \bibinfo{journal}{IEEE J. Quantum Electron.} \textbf{\bibinfo{volume}{28}},
  \bibinfo{pages}{93} (\bibinfo{year}{1992}).

\end{thebibliography}
 \end{document}

\title{Feedback-Induced Steady-State Light Bunching Above the Lasing Threshold}

\author{Franz Schulze}
\affiliation{Institut f\"ur Theoretische Physik, Nichtlineare Optik und Quantenelektronik, Technische Universit\"at Berlin, Hardenbergstra\ss e 36, 10623 Berlin, Germany}
\email[]{schulze@itp.tu-berlin.de}
\author{Benjamin Lingnau}
\affiliation{Institut f\"ur Theoretische Physik, Nichtlineare Dynamik und Kontrolle, Technische Universit\"at Berlin, Hardenbergstra\ss e 36, 10623 Berlin, Germany}
\author{Sven Moritz Hein}
\author{Alexander Carmele}
\affiliation{Institut f\"ur Theoretische Physik, Nichtlineare Optik und Quantenelektronik, Technische Universit\"at Berlin, Hardenbergstra\ss e 36, 10623 Berlin, Germany}
\author{Eckehard Sch\"{o}ll}
\author{Kathy L\"{u}dge}
\affiliation{Institut f\"ur Theoretische Physik, Nichtlineare Dynamik und Kontrolle, Technische Universit\"at Berlin, Hardenbergstra\ss e 36, 10623 Berlin, Germany}
\author{Andreas Knorr}
\affiliation{Institut f\"ur Theoretische Physik, Nichtlineare Optik und Quantenelektronik, Technische Universit\"at Berlin, Hardenbergstra\ss e 36, 10623 Berlin, Germany}

\date{\today}

\pacs{42.55.Sa, 42.50.Ar, 42.65.Sf, 42.55.Px}

\maketitle

\paragraph{Full quantum-optical set of dynamical equations}

\begin{table}[htb]
\begin{center}
\begin{tabular}{cc|cc}
\hline
\hline
Parameter & Value & Parameter & Value
\\
\hline
$\beta$ 			& $1 \times 10^{-04}$	& $\taumeinsrelc$ 		& $\frac{1}{1000}\,\text{fs}^{-1}$\\
$S = e^{- \kappaext \tau}$ 	& $0.5$			& $\taumeinsrelv$		& $\frac{1}{500}\,\text{fs}^{-1}$\\
$\NQD$ 				& $2000$		& $\taumeinsspe$, $\taumeinsspg$& $\frac{1}{50000}\,\text{fs}^{-1}$\\
$\tau$				& $90\,\text{ps}$	& $\gammaPD$			& $\frac{1.36\,\text{meV}}{\hbar}$\\
$\kappa^{-1},\kappah^{-1}$	& $22\,\text{ps}$	& 				& \\
\hline
\hline
\end{tabular}
\end{center}
\caption{Simulation parameters\label{table_parameters}}
\end{table}

\appendix

\begin{table}[htb]
\begin{center}
\begin{tabular}{c c|c c}
\hline
\hline
Symbol & Expectation Value & Symbol & Expectation Value
\\
\hline
$\fc$ & $\plfc$& $\cdagc$ & $\plcdagc$ \\
$\fv$ & $\plfv$& $\ddagqc$ & $\plddagqc$ \\
$\fce$ & $\plfce$ & $\ddagqdqs$ & $\plddagqdqs$ \\
\cline{3-4}
$\fve$ & $\plfve$ & $\avaccdag$ & $\plavaccdag$ \\
\cline{1-2}
$\cfccdagc$ & $\plcfccdagc$  & $\avacddagq$ & $\plavacddagq$ \\
\cline{3-4}
$\cfvcdagc$ & $\plcfvcdagc$  & $\cavaccdagcdagc$ & $\plcavaccdagcdagc$ \\
$\cfcddagqc$ & $\plcfcddagqc$ & $\cavaccdagcdagdq$ & $\plcavaccdagcdagdq$ \\
$\cfcddagqdqs$ & $\plcfcddagqdqs$  & $\cavacddagqcdagc$ & $\plcavacddagqcdagc$ \\
$\cfvddagqc$ & $\plcfvddagqc$ & $\cavacddagqcdagdqs$ & $\plcavacddagqcdagdqs$ \\
$\cfvddagqdqs$ & $\plcfvddagqdqs$ & $\cavacddagqddagqsc$ & $\plcavacddagqddagqsc$ \\
\hline
$\ccdagcdagcc$ & $\plccdagcdagcc$ & $\Delta^{\mathrm{n,k}}_{\mathrm{m,l}}$ & $(\omega_n - \omega_m + \omega_k - \omega_l)$ \\
$\cddagqcdagcc$ & $\plcddagqcdagcc$ & $\GAMMAextPD$ & $(\kappaext + \gammaPD)$ \\
$\cddagqcdagcdqs$ & $\plcddagqcdagcdqs$ & $\GAMMAzwextPD$ & $(2 \kappaext + \gammaPD)$ \\
$\cddagqsddagqcc$ & $\plcddagqsddagqcc$ & & \\
\hline
\hline
\end{tabular}
\end{center}
\caption{Notation\label{tab_abbreviations}}
\end{table}

\begin{align}
\label{eqs_cdagc}
\dt \cdagc			= &- 2 \cimag{\M \avaccdag 
					} \NQD
 				  + 2 \cimag{\sumq \Gsq \ddagqc }
\end{align}
\begin{align}
\label{eqs_rates}
\ratevps			=&  \fv  (1- \fve)  \taumeinsrelv
~~~\ratecps 			=  (1- \fc)  \fce  \taumeinsrelc
\end{align}
\begin{align} 
\label{eqs_fv}
\dt \fv				= &- 2\cimag{ \M 
					\avaccdag}
				  + (1-\beta)  \fc (1-\fv)  \taumeinsspg
				  - \ratevps
\end{align}
\begin{align}
\label{eqs_fc}
\dt \fc				= &+ 2 \cimag{ \M
					\avaccdag }
				  - (1-\beta) \fc (1-\fv) \taumeinsspg
				  + \ratecps
\end{align}
\begin{align}
\label{eqs_fve}
\dt \fve			= &- \taumeinspump  (\fve - \fce)
				  + (1- \fve)  \fce  \taumeinsspe
				  + \ratevps
\end{align}
\begin{align}
\label{eqs_fce}
\dt \fce			= \taumeinspump  (\fve - \fce)
				  - (1- \fve)  \fce  \taumeinsspe
				  - \ratecps
\end{align}
\begin{align}
\label{eqs_avaccdag}
(\dt &+ \gammaPD + \I\DELTAcvnull) \avaccdag	= 
\\
&				  - \I \Ms 
				  [
					\fc (1-\fv)
				  +
					\cfccdagc
				  -
					\cfvcdagc
				  +
					(\fc- \fv)  \cdagc
				  ]
				  \nonumber
\\				  
&				  - \I \sumq \Gsq \avacddagq
				  \nonumber
\end{align}
\begin{align}
\label{eqs_cfccdagc}
\dt  \cfccdagc		= 
				  &+ \cimag{
				  2\M 
					\cavaccdagcdagc
				  + 2\M 
					\avaccdag  ( \cdagc + \fc)
				  }
\\
				  &+ 2\cimag{\sumq \Gsq \cfcddagqc}
				  \nonumber
\end{align}
\begin{align}
\label{eqs_cfvcdagc}
\dt  \cfvcdagc			= 
				  &- \cimag{ 
				  + 2\M 
					\cavaccdagcdagc
				  + 2\M 
					\avaccdag  [ \cdagc + (1-\fv) ]
				  }
\\
				  &+ 2\cimag{\sumq \Gsq \cfvddagqc}
				  \nonumber
\end{align}
\begin{align}
\label{eqs_ccdagcdagcc}
\dt  \ccdagcdagcc		=  
				  &- 4 \cimag{ \M 
					\cavaccdagcdagc }  \NQD
				  + 4  \cimag{\sumq \Gsq \cddagqcdagcc}
\end{align}
\begin{align}
\label{eqs_cavaccdagcdagc}
(\dt &+ \gammaPD + \I\DELTAcvnull)  \cavaccdagcdagc		= 
				  + 2\I \M \fc  \cfvcdagc
\\
				  &- \I \Ms 
				  [
					(\fc- \fv)   \ccdagcdagcc
				  + 2
					\cfccdagc \left[(1-\fv) + \cdagc\right]
				  - 2
					\cfvcdagc  \cdagc
				  ]
\nonumber
\\
 				  &- 2\I \avaccdag
				  (
				  \M 
					\avaccdag
				  - \Ms 
					 \avaccdag^*
				  )
\nonumber
				  + \I \sumq \Gq \cavaccdagcdagdq
				  - 2\I \sumq \Gsq \cavacddagqcdagc
\nonumber
\end{align}
\begin{align} 
\label{eqs_ddagqc}
    (\dt &+ \kappaext - \I\DELTAomqnull) \ddagqc			=
\\
    					&+ \I \M 
					\avacddagq  \NQD
    				    - \I \Gq 
					\cdagc 
				    + \I \sumqs \Gqs \ddagqdqs
					\nonumber
\end{align}
\begin{align}
\label{eqs_avacddagq}
   (\dt &+ \GAMMAextPD + \I\DELTAcvq) \avacddagq		=   
\\
				    &- \I \Ms 
				    [
					(\fc-\fv) \ddagqc
				    +
					\cfcddagqc
				    -
					\cfvddagqc
				    ]
					\nonumber
  				    - \I \Gq 
					\avaccdag 
					\nonumber
\end{align}
\begin{align}
\label{eqs_cfcddagqc}
    (\dt &+ \kappaext - \I \DELTAomqnull) \cfcddagqc		=   
\\
				    &- \I \M 
				    (
				\avacddagq \cdagc
				    +
					\cavacddagqcdagc
				    +
					\avacddagq \fc
				    )
				    + \I \Ms 
				    (
					\cavaccdagcdagdq^*
				    +
					\avaccdag^* \ddagqc
				    )
					\nonumber
\\
	  			    &- \I \Gq 
					\cfccdagc 
				    + \I \sumqs \Gqs \cfcddagqdqs
				\nonumber
\end{align}
\begin{align}
\label{eqs_cfvddagqc}
    (\dt &+ \kappaext - \I\DELTAomqnull) \cfvddagqc		=   
\\
				    &+ \I \M 
				    [
					\avacddagq
					(1 - 
					\fv
				       + \cdagc )
				    +
					\cavacddagqcdagc
				    ]
					\nonumber
\\
				    &- \I \Ms 
				    (
					\avaccdag^* \ddagqc
				    +
					\cavaccdagcdagdq^*
				    )
					\nonumber
			            - \I \Gq 
					\cfvcdagc 
				    + \I \sumqs \Gqs \cfvddagqdqs
				    \nonumber
\end{align}
\begin{align}
\label{eqs_cddagqcdagcc}
    (\dt &+ \kappaext - \I\DELTAomqnull) \cddagqcdagcc		= + \I( 2 \M 
					\cavacddagqcdagc
				    - \Ms 
					\cavaccdagcdagdq^*) \NQD
\\					
				   &+ \I 2\sumqs \Gqs \cddagqcdagcdqs
				    \nonumber
				    - \I \sumqs \Gsqs \cddagqsddagqcc
  				   - \I \Gq 
					\ccdagcdagcc 
					\nonumber
\end{align}
\begin{align}
\label{eqs_cavacddagqcdagc}
    (\dt &+ \GAMMAextPD - \I \DELTAomqnull) \cavacddagqcdagc	=   - (2\I \M 
					\avaccdag
				    - \I \Ms 
					 \avaccdag^*)\avacddagq
\\
				    &- \I \Ms[ 
					\cfcddagqc (1
					+ \cdagc)
				    -
					\cfvddagqc \cdagc
					\nonumber
				    - (
					\cfvcdagc
				    -
					\cfccdagc) \ddagqc
				    ]
					\nonumber
\\
				    &+ \I \sumqs \Gqs \cavacddagqcdagdqs
				    - \I \sumqs \Gsqs \cavacddagqddagqsc
				    \nonumber
				    - \I \Gq 
					\cavaccdagcdagc 
					\nonumber
\end{align}
\begin{align}
\label{eqs_cavaccdagcdagdq}
(\dt &+ \GAMMAextPD + \I \DELTAomcvnullnullq) \cavaccdagcdagdq	= 
				    - 2\I \Ms 
					\conjcfcddagqc( 1
			+ \cdagc )
\\
				    &+ 2\I \Ms 
				    (
					\conjcfvddagqc \cdagc
				    +
					\avacddagq^* \avaccdag
				    )
					\nonumber
					+ \I \Gsq 
					\cavaccdagcdagc 
    				    - 2\I \sumqs \Gsqs \cavacddagqscdagdq
				    \nonumber
\end{align}
\begin{align}
\label{eqs_ddagqdqs}
(\dt + 2 \kappaext - \I\DELTAomqomqs) \ddagqdqs		=   	    &+ \I  \Gsqs 
					\ddagqc 
 				    - \I  \Gq 
					\ddagqsc^* 
\end{align}
\begin{align}
\label{eqs_cddagqcdagcdqs}
(\dt &+ 2 \kappaext 
+ 2 \kappah
- \I \DELTAomqomqs) \cddagqcdagcdqs	=   
\\
			    &+ \I(  \M 
					\cavacddagqcdagdqs
			    -
				\Ms 
					\conjcavacddagqscdagdq) \NQD
					\nonumber
			    - \I  \Gq 
					\conjcddagqscdagcc 
			    + \I  \Gsqs 
					\cddagqcdagcc 
\end{align}
\begin{align}
\label{eqs_cddagqsddagqcc}
(\dt &+ 2 \kappaext
+ 2 \kappah
- \I \DELTAomqomqsnullnull) \cddagqsddagqcc	=   
\\
 				    &+ 2 \I \M 
					\cavacddagqddagqsc \NQD
					\nonumber
				    - \I  \Gqs 
					\cddagqcdagcc 
 				    - \I \Gq 
					\cddagqscdagcc 
\end{align}
\begin{align}
\label{eqs_cavacddagqcdagdqs}
(\dt &+ \GAMMAzwextPD
+ \kappah
+ \I \DELTAomcvomqnullomqs) \cavacddagqcdagdqs	=
\\
 				    &- \I \Ms 
				    [
					\cfcddagqdqs(1
					+\cdagc)
					\nonumber
				    +(
					\fc
				   -
					\fv) \cddagqcdagcdqs
				   ]
					\nonumber
\\
 				    &- \I \Ms 
				    [
					(\conjcfcddagqsc
					- \conjcfvddagqsc) \ddagqc
					\nonumber
				    -
					\cfvddagqdqs \cdagc
					\nonumber
				    -
					\avacddagqs^* \avacddagq
				   ]
					\nonumber
\\
				    &- \I  \Gq 
					\cavaccdagcdagdqs 
 				    + \I \Gsqs 
					\cavacddagqcdagc 
				    \nonumber
\end{align}
\begin{align}
\label{eqs_cavacddagqddagqsc}
(\dt &+ \GAMMAzwextPD 
+ \kappah
+ \I \DELTAomcvomnullomqomqs
)  \cavacddagqddagqsc	= 
\\
			    	    &- \I  \Ms
				    [
					(\fc 
				-	\fv)
					\cddagqsddagqcc
				    + (
					\fc 
					- \fv )
					\ddagqsc \ddagqc
				   ]
					\nonumber
\\
 				    &- \I  \Ms 
				    [
				       (\cfcddagqsc
				-	\cfvddagqsc) \ddagqc
					\nonumber
				    +
				       (\cfcddagqc
				-       \cfvddagqc ) \ddagqsc
				    ]
					\nonumber
\\
 				    &- 2 \I  \M 
					\avacddagq \avacddagqs
					\nonumber
				    - \I  \Gqs 
					\cavacddagqcdagc 
 				    - \I  \Gq 
					\cavacddagqscdagc 
					\nonumber
\end{align}
\begin{align}
\label{eqs_cfvddagqdqs}
(\dt &+ 2 \kappaext - \I\DELTAomqomqs) \cfvddagqdqs            =	   
\\
 				    &+ \I \M 
				    (
					\cavacddagqcdagdqs
				    +
					\avacddagq \ddagqsc^*
				    )
				    - \I \Ms 
				    (
					\conjcavacddagqscdagdq
				    +
					\avacddagqs^* \ddagqc
				    )
					\nonumber
\\				    
				    &- \I \Gq 
					\conjcfvddagqsc 
 				    + \I \Gsqs 
					\cfvddagqc 
					\nonumber
\end{align}
\begin{align}
\label{eqs_cfcddagqdqs}
(\dt &+ 2 \kappaext - \I\DELTAomqomqs) \cfcddagqdqs	=   
\\
 				    &- \I \M 
				    (
					\cavacddagqcdagdqs
				    +
					\avacddagq \ddagqsc^*
				    )
				    + \I \Ms 
				    (
					\conjcavacddagqscdagdq
				    +
					\avacddagqs^* \ddagqc
				    )
					\nonumber
\\
				    &- \I  \Gq 
					\conjcfcddagqsc
 				    + \I \Gsqs 
					\cfcddagqc 
					\nonumber
\end{align}